# Simultaneous measurement of the exchange parameter and saturation magnetization using propagating spin waves

Grant A. Riley[1,2], Justin M. Shaw[2], Thomas J. Silva[2], Hans T. Nembach[2,3]*

[1]Center for Memory and Recording Research, University of California-San Diego, La Jolla, CA 92093 USA

[2]Quantum Electromagnetics Division, National Institute of Standards and Technology, Boulder, Colorado 80305, USA

[3]Department of Physics, University of Colorado, Boulder, Colorado 80309, USA

The exchange interaction in ferromagnetic ultra-thin films is a critical parameter in magnetization-based storage and logic devices, yet the accurate measurement of it remains a challenge. While a variety of approaches are currently used to determine the exchange parameter, each has its limitations, and good agreement among them has not been achieved. To date, neutron scattering, magnetometry, Brillouin light scattering, spin-torque ferromagnetic resonance spectroscopy, and Kerr microscopy have all been used to determine the exchange parameter. Here we present a novel method that exploits the wavevector selectivity of Brillouin light scattering to measure the spin wave dispersion in both the backward volume and Damon-Eshbach orientations. The exchange, saturation magnetization, and magnetic thickness are then determined by a simultaneous fit of both dispersion branches with general spin wave theory without any prior knowledge of the thickness of a magnetic "dead layer". In this work, we demonstrate the strength of this technique for ultrathin metallic films, typical of those commonly used in industrial applications for magnetic random-access memory.

*Introduction* – The Heisenberg exchange coupling between neighboring spins gives rise to ferromagnetic order and is a crucial material parameter in data storage applications[1]. A variety of methods have been previously used to determine the exchange parameter, $A_{ex}$, including neutron scattering[2,3], temperature-dependent magnetometry[4], characterizing perpendicular standing spin waves (PSSWs) with either Brillouin light scattering (BLS) or ferromagnetic resonance (FMR)[5–8], the measurement of Damon-Eschbach and backscattering mode-dispersion with BLS[9,10], and spin-torque FMR measurements of localized spin wave modes in devices[11,12]. However, additional challenges arise for some of these techniques when extended to films thinner than 10 nm.[13]

Neutron scattering measurements are considered the "gold standard" for exchange measurements because they probe magnons with large wavevectors, where exchange is the dominant energy contribution. However, thick films are required due to the low scattering cross-section. Moreover, neutron scattering requires access to a large-scale facility, impeding routine measurements. Measurements based on PSSWs require detailed information about the pinning conditions of the magnetic moments at interfaces, which is often difficult to determine. Exchange measurements have also been demonstrated with spin-torque FMR,[11,12] by measuring the characteristic frequencies of localized spin waves in the free layer of a magnetic tunnel junction to determine the exchange stiffness. Not only does this method require fabrication of a full device, the characteristic frequencies of patterned magnetic microstructures are also inherently more complex than for extended thin films and can be strongly affected by defects and pinning conditions at the interfaces and edges of the microstructure.[14] In this work, we also explore the previously used approach of measuring the spin wave frequency at various angles of incidence with the external field perpendicular to the plane of incidence. Briefly, this approach results in an unambiguous loss of accuracy that becomes increasingly significant in thinner films.

*Corresponding author: hans.nembach@nist.gov

The exchange parameter has also been measured with temperature-dependent magnetometry by use of the Bloch $T^{3/2}$ law. This method provides an accurate measure of the saturation magnetization $M_S$, if the magnetic thickness $t$ of the film is accurately known. However, many films contain a magnetically "dead layer", which must be ascertained by the measurement of an entire thickness series. This is particularly important in the ultrathin limit (<2 nm), where the dead layer constitutes a significant portion of the total thin-film volume[15]. Furthermore, there has been a wide variety of phenomenological approaches to fitting high-temperature $M_S$ values including a range of possible exponents for temperature dependence[16–20]. Also, possible deviations from the Bloch law with reduced dimensionality have been discussed[20]. The method to convert the low-temperature exchange parameter to room temperature and higher, which is often of most interest for technological applications, relies on mean-field theory models that require additional assumptions[21]. These challenges highlight the need for a direct measure of exchange at smaller wavevectors and at room temperature that is non-destructive, readily accessible for the purposes of technology development, and even applicable for wafer-level measurements on the manufacturing scale.

In this paper, we present a method to simultaneously quantify the exchange parameter, saturation magnetization, and magnetic thickness without prior knowledge of the dead layer thickness from BLS measurements of both the Damon-Eshbach *and* the backward volume spin wave modes by varying both the optical angle of incidence and the scattering plane with respect to the applied magnetic field direction. This approach allows us to extract $A_{ex}$, $M_S$ and $t$ from a single set of measurements by simultaneously fitting the data to the well-established spin wave dispersion relations. There are several factors that make such an approach desirable. First, $A_{ex}$, $M_S$, and $t$ are extracted with no additional assumptions, including the sample magnetic thickness. As such, this method offers the possibility of in-line wafer-level elevated-temperature inspection of $A_{ex}$ and $M_S$ for both product development and manufacturing applications.

*Experiment* – We deposited a series of Ta/Cu/Co$_{90}$Fe$_{10}$($t$)/Cu/Ta films on quarter wafers of thermally oxidized silicon substrates using sputtering, with physical thicknesses of $t_P$ = 1.4, 3.0, and 14.0 nm. The deposition rates were calibrated by X-ray reflectometry. The thickness series allows determination of the thickness, $t_D$, of the magnetically dead region near the interfaces, which is required for the determination of $M_s$ from SQUID (Superconducting Quantum Interference Device) magnetometry. An additional 6 nm film was deposited to increase the accuracy of the dead layer determination. The ferromagnetic layer is sandwiched between Cu layers in our sample structure to minimize the dead layer thickness, as well as to minimize any effects of spin-pumping on the linewidth, which would reduce the signal strength. Each wafer was diced into identical 6 mm x 6 mm chips for the SQUID measurements and 3 mm x 3 mm chips for the BLS measurements taken from the center of the wafer to ensure uniformity. The samples were measured using ferromagnetic resonance (FMR), SQUID magnetometry, and Brillouin light scattering (BLS).

As discussed earlier, BLS has been used to probe spin waves in only one field orientation, namely the Damon-Eschbach (DE) configuration, where the wavevector is perpendicular to the magnetization. We also measure backward volume (BV) spin waves, another common configuration where the wavevector is parallel with the magnetization, and where the phase and group velocities point in opposite directions in a certain wavevector range. To measure both DE and BV spin waves, we designed a sample holder that incorporates two permanent magnets to apply an in-plane field in directions parallel and perpendicular to the scattering plane. Further detail on the BLS measurement configuration is provided in the *Supplemental Information*.

The data for our samples are shown in Fig. 1. These two branches of the dispersion manifold are simultaneously fit by the approximate dispersion relation[22]:

$$f(k, H, \phi, t) = \frac{\mu_0 \mu_B g}{h} \sqrt{\left(H - H_k + \frac{2A_{ex}}{\mu_0 M_S}k^2 + \mathcal{N} M_S \left(\frac{1-e^{-kt}}{kt}\right)\right)\left(H + \frac{2A_{ex}}{\mu_0 M_S}k^2 + \mathcal{N} M_S \left(1 - \frac{1-e^{-kt}}{kt}\right)\sin^2(\phi)\right)} \quad (1)$$

where $f$ is the spin wave frequency, $g$ is the spectroscopic splitting factor, $H$ is the external field, $H_k$ is the out-of-plane anisotropy, $A_{ex}$ is the exchange parameter at room temperature, $M_S$ is the saturation magnetization at room temperature, $k$ is the in-plane spin wave wavenumber, $t$ is the magnetic thickness ($t = t_p - t_D$), and $\phi$ is the angle between the spin wave wavevector and the magnetization, with $sin^2(\phi) = 1$ and $sin^2(\phi) = 0$ for the DE and BV geometries, respectively. $\mathcal{N}$ is an approximate perpendicular demagnetization factor for ultrathin films, given by $\mathcal{N} \cong 1 - 0.2338/n$, where $n$ is the number of monolayers[23]. Typically, the demagnetization factor is calculated using a continuum model; however, that model does not apply for samples with a few atomic layers due to a dipolar field that varies with the number of monolayers.[24] The fractions appearing in the terms with the demagnetizing factor are also an approximation for thin films. By fitting our data with eq. (1), $A_{ex}$, $M_S$, and $t$ are extracted at room temperature as the sole fitting parameters.

The FMR measurements were performed with a vector network analyzer (VNA) to characterize the spectroscopic splitting factor *g*, the effective magnetization *M*<sub>eff</sub>, and the out-of-plane magnetic anisotropy $H_k^\perp$. The films were measured over a broad frequency range of 10 GHz to 35 GHz, and with the static magnetic field applied both within and perpendicular to the film plane. Asymptotic analysis was used for accurate determination of *g* and *M*<sub>eff</sub> as described in Ref. 25. The results of the FMR measurements are summarized in the *Supplemental Information*.

To compare the BLS-based method with a previously established method, temperature dependent SQUID magnetometry measurements were also performed. The thickness- and temperature-dependent saturation magnetic moment $m_s(T, t)$, where $t$ is the magnetic thickness, was determined from measurements of the hysteresis curve at temperatures ranging from 20 K to 300 K. After accounting for the diamagnetic contribution of the substrate, a linear fit of $m_s(T = 0 \text{ K})$ vs $t_P$ was performed, where the x-intercept is the magnetic dead layer $t_D$. Having determined $t$, along with the macroscopic sample dimensions, measured using an optical microscope, the total volume of the ferromagnet was determined. The measured magnetic moment $m_S(T, t)$ was then fit using the Bloch law[26]:

$$m_S(T, t) = m_S(T = 0 \text{ K}, t)\left(1 - \frac{g \mu_B \eta}{M_S(T=0 \text{ K}, t)}\left(\frac{k_B T}{D_{spin}(T=0 \text{ K}, t)}\right)^{3/2}\right) \quad (2)$$

A detailed discussion of the how to obtain a room temperature measure of $A_{ex}$ from the Bloch law fits is provided in the *Supplemental Information*.

**Results** -The FMR measurement results, summarized in Table S1 in the SI, shows that, $M_{eff}$ increases monotonically with film thickness, while $H_k$, calculated using measured values of $M_S(T = 300 \text{ K}, t)$, decreases monotonically with film thickness, as expected in the case of interfacial anisotropy. Due to our measurement geometry, $g^\parallel$ is the relevant parameter in the dispersion relation for BLS. $g^\parallel$ varied by less than 1 % between the four samples, with a weighted mean of 2.173(2).

The BLS results for the three films in both the BV and DE geometries are shown in Fig. 1. The balance between the magnetostatic energy of the dipolar-interaction and the exchange interaction leads to the inflection of curvature over the range of thicknesses chosen. The measured dispersion for the thinnest film has a noticeably parabolic curvature due to the quadratic dependence of the frequency on the wavevector that results in a lower uncertainty in the measured exchange parameter. As the thickness increases, the dipolar energy also increases leading to a compensation of the quadratic term that results in a larger uncertainty as the film thickness increases. It should be noted that no higher order thickness modes are visible within the BLS spectra because they have frequencies larger than the free spectral range of the interferometer.

The results of the Bloch law fit given in equation (2) for each sample are show in Figure 2. The magnetic moment at saturation was found to be $m_S(T = 0 \text{ } K, t) = 72.068(32)$, and $155.64(11)$, and $739.76(12) \text{ n Am}^2$ and the spin-stiffness constant was found to be $D_{spin}(T = 0 \text{ K}, t)/10^{-40} = 6.561(62), 6.20(11)$, and $5.931(90) \text{ Jm}^2$ for the 1.4, 3, and 14 nm films respectively. These values were used to determine $\mu_0 M_S(T_{RT}, t)$ and $A_{ex}(T_{RT}, t)$, given in Table 1, as described above. The thickness dependence of the magnetic moment is shown in the *Supplemental Information* where the *x*-intercept indicates a dead layer of $0.04(1)$ nm at 0 K.

Table 1 shows the results from both the SQUID data and the dispersion curve fits to the BLS data with $A_{ex}$, $M_S$ and *t* as fitting parameters. The values in brackets are obtained from a fit where the film thickness was determined from the thickness series and was not a fitting parameter. The exchange parameter for the three films measured with BLS have overlapping error bars with their weighted mean value of $A_{ex}(T_{RT}) = 21.020(50) \text{ pJ/m}$ which is within 3 % of the SQUID-measured weighted mean value of $A_{ex}(T_{RT}) = 21.488(52) \text{ pJ/m}$. The BLS-measured values of exchange for the 14 nm sample compares well with the previously obtained result of 22.8 pJ/m for bulk $Co_{92}Fe_8$[13], measured with neutron scattering (here we have used our measured values for *g* and $M_S$ for the 14 nm film to convert to $A_{ex}$).

**Discussion** – Constraining the fit to the BLS data to use $t_D$ from the thickness series, instead of the magnetic thickness being a fitting parameter, did not have a significant impact on the extracted parameter agreement. Notably, the measured values of $A_{ex}$ agree for all three films using both techniques within one standard error. The $M_S$ values do not agree within two standard errors between SQUID and BLS measurements for the 1.4 and 3 nm films but agree for the 14 nm film to within one standard error. We note that the reported uncertainty for both techniques originate from fitting, not repeated measurements. While the error bars do not overlap for the thinner films, the measurements agree to within 6 % for the 1.4 nm film and less than 2 % for the 3 nm film indicating that the BLS technique provides a reliable estimate for the saturation magnetization. Discrepancy between the two techniques is possibly due to the assumptions required in the Bloch law fitting procedure, namely a proper determination of the density of states, the choice of the exponent and temperature range. In addition, the conversion of $A_{ex}$ at low temperature to a room temperature value introduces additional uncertainty. The exchange parameter deviates significantly over this temperature range, however mean field theory used for the conversion faces difficulty closer to the Curie temperature. Moreover, the exchange value in the Bloch law is assumed to be temperature independent. Here, we used the exchange value averaged over the temperature range of the fits. This highlights an important aspect of the BLS technique, that $A_{ex}$ at room temperature is a direct fit parameter from the dispersion theory, which requires only previously examined assumptions, namely the approximate demagnetization factor and the use of an approximate closed form solution to the dispersion relation.

To better understand the necessity to measure both the DE and BV branches of the dispersion curves, independent fits of the same experimental data were performed using the DE or BV branch alone. These fits along with Monte Carlo simulations are provided in the *Supplementary Information*. Fits to the BV branch alone are particularly inaccurate, with measured values of both $M_S$ and $A_{ex}$ deviating from the full fit by more than two standard errors for all three films. Furthermore, the magnitude of the standard error increases by up to a factor of 17 for the partial fit indicating a significant reduction in precision. Fits to the DE branch alone are more precise with an increase up to a factor of 5 in relative error from the full fit. However, the measured values deviate by more than two standard errors for all three films. The Monte Carlo simulations confirm that the reduction in precision from fitting only one branch is not unique to the measured data set. In the full fit, there is a general decrease in accuracy with increase in film thickness that may be due to higher order effects not accounted for in the dispersion relations, such as a change of the saturation magnetization or the exchange parameter through the thickness of the ferromagnetic layer. It has been reported that both quantities are thickness dependent for thin films.[27]

Measurements of the field gradients of the permanent magnets necessitated the use of fiducial marks for precise positioning of the BLS probe on the sample. The BLS measurement technique could be further improved by replacing the permanent magnets with an electromagnet with pole pieces that lead to increased field uniformity and allow for a wide range of optical access and by reducing the focused spot size of the BLS with a modestly higher power focusing lens – using a lens with too much power would result in a reduction of wavevector resolution which might have a deleterious effect the dispersion curve fit results. While we chose a roughly uniform distribution of angles, one could also consider optimizing the measurement time by mapping out regions of the full dispersion manifold, guided by Bayesian statistics.

**Conclusion** – We have demonstrated a novel technique for measuring the exchange parameter in ultrathin $Co_{90}Fe_{10}$ films using Brillouin light scattering to measure propagating spin waves. In a 1.4 nm thick film we measured 20.9(6) pJ/m, which agrees with our SQUID measured value of 21.5(1) pJ/m within one standard error and is close to a previous experimental result of 22.8 pJ/m for *bulk* $Co_{92}Fe_8$[13]. Using our technique, we obtain the exchange parameter from a fit with established spin wave dispersion theory and simultaneously obtain agreement within 6% of SQUID measured values of the saturation magnetization for all film thicknesses measured. This novel approach opens new possibilities for accurate measurements of room temperature exchange, saturation magnetization and effective magnetic thickness in ultrathin films and could be utilized for non-destructive inline measurements on the wafer scale. This technique is applicable to even thinner films than used here and can possibly be multiplexed– utilizing multiple laser beams in parallel, each probing spinwaves at a different wavevector, to dramatically reduce the data-acquisition time. This technique will lead to further understanding the physics behind the exchange interaction and its impact in devices.

**Acknowledgement** – The authors would like to thank Monika Arora for growing the thickness series of films used in this study. GR acknowledges support from DOE award No. DE-SC0018237.

| $t_N(nm)$ | $t = t_N - t_D (nm)$ [BLS] | $A_{ex}(pJ/m)$ [BLS] | $\mu_0 M_S(T)$ [BLS] | $A_{ex}(pJ/m)$ [SQUID] | $\mu_0 M_S(T)$ [SQUID] |
|---|---|---|---|---|---|
| 1.4 | 1.36(3) {1.36(1)} | 20.9(6) {20.9(5)} | 1.881(2) {1.881(1)} | 21.5(1) | 1.769(25) |
| 3 | 3.30(7) {2.96(1)} | 21.3(1.3) {20.4(2.6)} | 1.769(7) {1.792(12)} | 21.4(2) | 1.804(48) |
| 14 | 13.93(25) {13.96(1)} | 22.4(2.4) {22.6(1.5)} | 1.856(13) {1.854(5)} | 21.6(1) | 1.869(43) |

*Table 1: Results for magnetic thickness, t, the exchange parameter, $A_{ex}$, and the saturation magnetization, $M_s$, from dispersion curve fits to BLS measurements performed in the DE and BV geometries and Bloch-law fits for magnetometry measurements made with SQUID. All values are reported at room temperature. The bracketed values are from fits where magnetic thickness was not a fitting parameter. Instead, the magnetic thickness determined by magnetometry of the entire sample series was utilized. The $M_s$ measured by SQUID is provided here for comparison.*

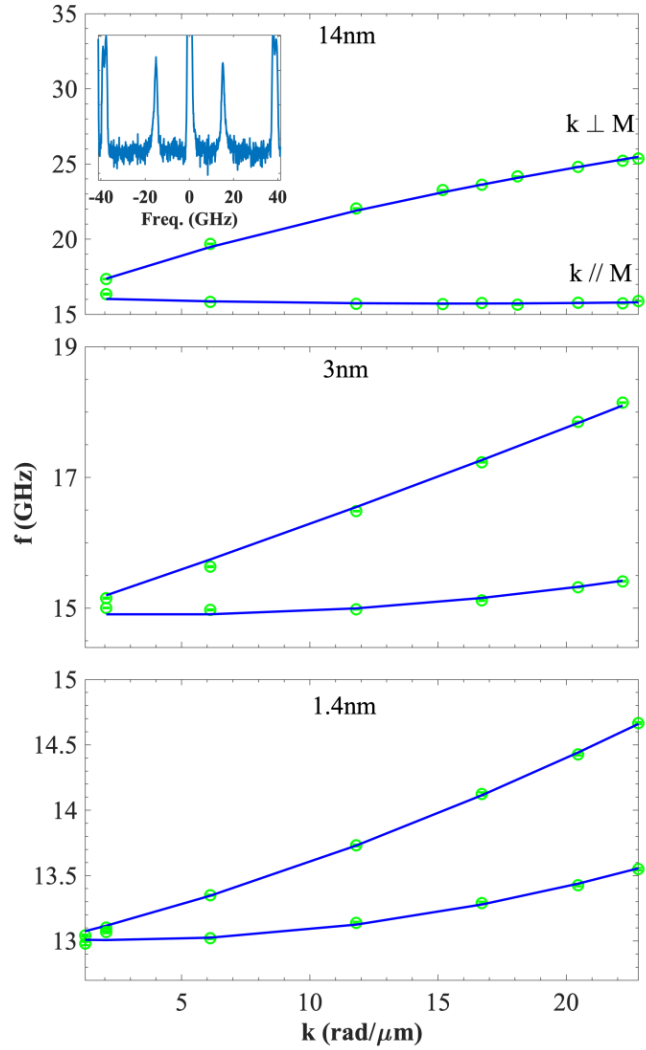

Figure 1: Frequency of propagating spin waves in both the backward volume ($k \parallel M$) and Damon-Eschbach ($k \perp M$) geometries as measured by Brillouin light scattering. The scattering geometry determines the range of wavevectors. The error bars are included but mostly obscured by the data markers. The external field was fixed at $\mu_0 H = 0.15\ T$ for all of the measurements.

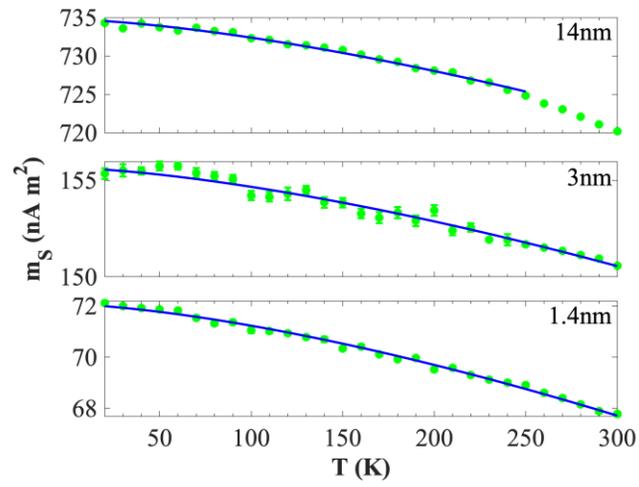

Figure 2: The magnetic moment at saturation as a function of temperature for all three films. Each data point corresponds to a hysteresis measurement at a fixed temperature. The Bloch law is used to fit the magnetic moment as a function of temperature to extract both $m_S(T = 0\ K)$ and $D_{spin}(T = 0\ K)$.